\documentclass{svjour3}                     % onecolumn (standard   format)
\smartqed  % flush right qed marks, e.g. at end of proof
\usepackage{graphicx}
\usepackage{epstopdf}
\usepackage{graphicx,graphics,color,epsfig}
\usepackage{mathptmx}      % use Times fonts if available on your TeX system
\usepackage{amsfonts}
\usepackage{amssymb}%
\begin{document}

\title{Entanglement Generation Between Two Mechanical Resonators in Two Optomechanical Cavities%\thanks{Grants or other notes
}
%\subtitle{Optomechanical Entanglement}

\titlerunning{Stationary Optomechanical Entanglement}        % if too long for running head

\author{Adel AL Rehaily         and
        Smail Bougouffa %etc.
}


\institute{A. Al Rehaily \at
              Department of Physics, Faculty of Science, Taibah University,
P.O.Box 30002, Madinah, Saudi Arabia \\
              \email{ad1399el@hotmail.com}           %  \\
           \and
           S. Bougouffa (Corresponding Author)\at
              Physics Department, College of Science, Imam Muhammad ibn Saud Islamic University (IMSIU), P.O. Box 90950, Riyadh 11623, Saudi Arabia\\
              \email{sbougouffa@hotmail.com or sbougouffa@imamu.edu.sa}
}

\date{Received: date / Accepted: date}
%The correct dates will be entered by the editor

\maketitle

\begin{abstract}
A standard model is suggested to explore correlation features of two spatially separated optomechanical cavities. The cavities are coupled through the photon-hopping process. In particular, we investigate the generation of entanglement between mechanical resonators in the strong coupling regime and the two cavities are assumed to be driven by a coherent laser field.
In order to quantify entanglement we use the logarithmic negativity. The analytical solutions are presented for the system in a parameter regime very close to the current experimental results. We show that in the presence of the photon hopping process between the cavities, the two mechanical resonators and the field modes can be entangled. This shows clearly that the entanglement can be transfer via radiation pressure of a photon hopping coupling from the intracavity photon-phonon entanglements to an inter-cavity photon-photon or phonon-phonon entanglement.

\keywords{Entanglement; Coupled Optomechanical Cavities; Optomechanics.}

\PACS{03.67.Bg \and 03.65.Ta \and 42.50.Wk \and 42.65.Yj \and 42.50.Pq }

\end{abstract}

\section{Introduction}
\label{intro}
Quantum correlations \cite{00}, as a foundation of quantum physics, play a crucial role in the source of quantum theory and also have large applications in quantum technology. On the other hand, the generation of quantum effects in various microscopic and macroscopic scales has received a great deal of attention \cite{01,01a,02}.  With the recent enlargement in laser cooling techniques, production of low-loss optical components and high-Q mechanical oscillators, it is now achievable to set up nanomechanical resonators, which may be controlled to a very high accuracy and can still reach the quantum stage of the oscillations \cite{02}. In particular, entanglement, which is a kind of quantum correlations, has been envisaged in various quantum systems \cite{03}. Furthermore, it is identified as one of the distinctions between classical and quantum worlds. However, it becomes a fundamental resource for various quantum algorithms such as quantum teleportation \cite{1,2}, quantum dense coding \cite{3}, quantum cryptography \cite{4}, and quantum computing \cite{5}.

Despite of a considerable development there still exist a number of unanswered concerns even though the evolutions in the recent years. In particular, the categorization and existence of entanglement for multipartite quantum systems are far from complete\cite{6,7}. Therefore, the vitality of entanglement demands two matters to be fronted: one is the successful generation of entanglement between two or more subsystems and the second is the problem of the degradation of entanglement with time due to decoherence which is the result of interaction with the environments. The enquiry of entanglement generation has been addressed several times and there are many experimentally realizable schemes for this purpose \cite{8,9}. So, once the entanglement creates between two or more subsystems is not very well protected from the environments. Since entanglement is a very breakable quantity, it may be totally destroyed when the quantum system interacts with the environment.

On the other hand, an interesting concern becomes recently very significant demonstrating how to generate macroscopic mechanical entanglement. The macroscopic entanglement may present specific facts for quantum phenomena \cite{10,11,12} and can explain the quantum-to-classical passage, as well as the limit between quantum and classical fields \cite{13}. Furthermore, the generation of quantum entanglement in a macroscopic mechanical system constitutes an interesting field of investigation and has attracted the attention of many researchers. Several schemes have been proposed to create quantum entanglement in diverse mechanical resonators \cite{14,15,16,17,18,19,20,21,22,23,24,25,25a,25b,25c,25d}.

In this way, cavity optomechanics \cite{27,28} can offer a normal basis to introduce a correlation between mechanical resonators since the cavity optomechanics explore a natural interaction between mechanical and optical degrees of freedom. This vital characteristic of opmechanical systems is at the heart of the motivation to investigate the macroscopic mechanical entanglement creation in a two-cavity optomechanical system\cite{29,30}.

Recently, it has shown that the possibility to generate non classical states of optical and mechanical modes of optical cavities distant from each other \cite{30a}.  Nevertheless, the degree of entanglement dynamics was not large. On the other hand, to increase the amount of entanglement, it has proposed an emplacement of optical parametric amplifiers inside two spatially separated coupled optomechanical systems \cite{30b}. However, with new advanced experiment results \cite{24,25}, we can generate an important amount of stationary entanglement between coupled optomechanical system without recourse to a complex situations.

In this work, we explore the macroscopic mechanical entanglement generation in two coupled cavity optomechanics. The proposed scheme is constituted of two coupled quantum cavity optomechanical systems. Each cavity  is a Fabry-Perot cavity composed of one moving end mirror, which is nano-or micro- mechanical vibration object. In such cavity, the field and mechanical modes are coupled.  We establish a correlation between the two field modes through a photon hopping interaction. This coupling process will serve to create an entanglement between the two mechanical modes. Recently, an analogue scheme is explored in the strong coupling regime and the deep resolved sideband regime, and it has been shown that the entanglement between the mechanical resonators can be generated \cite{31}. On the other hand, It has also been shown that the photon-phonon entanglement can be generated in the emission and scattering processes with single photon approximation \cite{31a}. In addition, it was treated as a bipartite one from the view point of single photon rather than photon modes.

Consequently, a question arises whether a driven system can also generate a significant entanglement between the two mechanical resonators without limitation to the single photon assumption. On the other hand, we are concerned with the behavior of the stationary  generated entanglement in terms of different parameters of the proposed scheme. In the following, we will investigate this point of view.

The rest of this paper is organized as follows.  We start in Sec.~\ref{sec:1} with the description of the two coupled cavity model and we present the Hamiltonian that governs the proposed scheme. We then derive in Sec.~\ref{sec:2}, the effective quantum Langevin equations (QLE) describing the dynamics of the system in the rotating wave approximation (RWA). We then employ the linearization technique to the equations of motion and get a set of coupled differential equations for the fluctuation operators, which then are solved for the steady state. In Sec.~\ref{sec:3} we present the steps of the method to calculate the covariant matrix in order to quantify the entanglement between different bipartite modes of the proposed scheme. In this context, we use the logarithmic negativity, which is a good and suitable measure for entanglement. In Sec.~\ref{sec:4}, we examine parameter ranges in which the predicted coherence and correlation effects could be obtained with the current experiments and we discuss our obtained results. Finally, in Sec.~\ref{sec:5}, we conclude our results.

\section{Model and Hamiltonian}
\label{sec:1}
\begin{figure}[h]
 \center{\includegraphics[width=0.70\linewidth,height=0.20\linewidth]{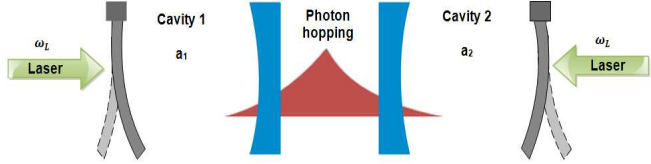}}
\caption{Schematic description of the system under study. Two separated optomechanical cavities, each driven by a laser.}\label{Fig1}
\end{figure}

We consider two spatially separated optomechanical cavities. Each cavity contents a fixed end mirror and a mechanical resonator. The cavity fields are coupled via a photon-hopping interaction. Therefore, each cavity field couples to the mechanical motion of the mechanical resonator through the radiation pressure interaction, see figure \ref{Fig1}. The field in each cavity is assumed to be a single field mode of frequency $\omega_{cj},(j=1,2)$.

The Hamiltonian of the hybrid system can be written as ($\hbar=1$)
\begin{eqnarray}
 \hat{H}&=&\sum_{j=1,2}\big[\omega_{cj} a_{j}^\dagger a_{j}
         + \frac{1}{2}\omega_{mj}(p_{j}^2+ q_{j}^2)
         - g_{j} a_{j}^\dagger a_{j} q_{j}\big]\nonumber\\
        &+&\sum_{j=1,2}i\big[E_j a_{j}^\dagger e^{-i\omega_{Lj}t}-E_j^{*} a_{j} e^{i\omega_{Lj}t}\big]
         -\xi \big( a_{1}^\dagger a_{2}+a_{2}^\dagger a_{1}\big),
\end{eqnarray}
In the first brackets, $ a_{j}^\dagger$ and $a_{j} $ are the creation and annihilation operators of the single cavity mode, which satisfy the commutation relation$([a_{j}^\dagger,a_{k}]=-\delta_{jk})$ with frequency $\omega_{cj} (j=1,2)$. $\omega_{mj}$ is the frequency of the mechanical oscillator, $q_{j}$ and $p_{j}$ are their dimensionless position and momentum operators, which satisfy the commutation relation $([q_{j},p_{k} ]=i \delta_{jk})$.
 The last term in this brackets describes the radiation pressure interaction with coupling strength $g_{j} = (\omega_{cj}/L_{j})\sqrt{ \hbar/(m_{j} \omega_{mj})}$, where $m_{j}$ is the mass of the mechanical mode and $L_{j}$ the rest length of the optomechanical cavity. The term in the second brackets represents the driving field with frequency $\omega_{Lj}$ and the amplitude $E_j$ which is expressed in terms of the input field power $P_{j}$ by $|E_j|=\sqrt{\frac{2P_{j}\kappa_{j}}{\hbar\omega_{Lj}}}$, where $\kappa_j$
is the decay rate of the cavity field. Without loss of generality, we assume that $\omega_{L1}=\omega_{L2}$. The last term describes the photon hopping coupling between the two cavity modes with a strength $\xi$. It is interesting to mention here that some previous investigations have considered multicavity opto-mechanical systems with one mechanical oscillator \cite{32,33} or two mechanical resonators in the deep-resolved-sideband regime \cite{31}. In the following, we are concerned in the dynamics of the system that can be determined by the quantum Langevin equations.

\section{Equations of Motion}
\label{sec:2}
An appropriate investigation of the problem requires including different effects. The main effect that to be taken in the analysis is the photon losses in the optical cavity that is characterized with the decay rate $\kappa_{j} $ and the loss of mechanical excitations, i.e. phonons, which is quantified by the energy dissipation rate $\gamma_{mj}=\omega_{mj}/Q_{mj}$  where $Q_{mj}$ is the mechanical quality factor.  The motion's equations can be deduced using Langevin equation. Thus the set of nonlinear Langevin equations for this system can be written in the interaction picture  with respect to $\hbar\omega_{Lj} a^{\dag}_{j}a_{j}$
\begin{eqnarray}
    \dot{q_{j}}&=&\omega_{mj} \ p_{j}, \nonumber\\
    \dot{p_{j}}&=&-\omega_{mj} \ q_{j} + g_{j} a_{j}^{\dag}a_{j}-\gamma_{mj} \ p_{j}+b^{in}_{j}, \nonumber\\
    \dot{a_{j}}&=&-(\kappa_{j}-i \Delta_{0j})a_{j} +i g_{j} a_{j} q_{j} +i \xi a_{k}+E_j +\sqrt{2\kappa_{j}} a^{in}_{j}.
\end{eqnarray}
These equations represent the equations of motion for the dimensionless position, momentum and photon annihilation operators, respectively. Here, $ \Delta_{0j} = \omega_{cj} - \omega_{Lj}$ is the cavity detunings, ($j,k=1,2$ with $k \neq j$).
We note that the nonzero correlation function for the vacuum input noise of the cavity $a_j^{in}$ and the Hermitian Brownian noise operator of the mechanical mode $b_j^{in}$ satisfy the following moments \cite{34}
\begin{equation}
\langle a_j^{in}(t) {a_j^{in}}^{\dagger}(t)\rangle=\delta(t-t'),
\end{equation}
\begin{equation}
\langle b_j^{in}(t) b_j^{in}(t')\rangle=\frac{\gamma_{mj}}{2\pi\omega_{mj}}
\int d\omega e^{-i\omega( t-t')}\omega
\big[\coth \left(  \frac{\hbar \omega}{2k_{B}T}\right)+1  \big],
\end{equation}
where $k_{B}$ is the Boltzmann constant and $T$ is the temperature of the reservoir of the mechanical resonator. It is clear that $b_j^{in}(t)$ does not describe a Markovian process. On the other hand, the quantum effects are reachable just by means of oscillators with a large mechanical quality factor $Q_{m_{j}}=\omega_m/\gamma_m \gg 1$. In this limit, we recover a Markovian process and $b_j^{in}$ will satisfy the following second moments
\begin{equation}
\frac{1}{2}\big\langle b_j^{in}(t) b_j^{in}(t')+b_j^{in}(t') b_j^{in}(t)\big\rangle \simeq\ \gamma_{mj}(2\bar{n}+1)\delta(t-t'),
\end{equation}
where $\bar{n} = (e^{\hbar \omega_{mj}/k_{B} T} - 1)^{-1} $ is the mean thermal excitation number at the frequency of the mechanical mode.

The steady state values for these nonlinear equations are given by
\begin{eqnarray}\label{5}
p^s_{j}&=&0,\nonumber\\
q^s_{j}&=&\frac{g_{j} \mid a^s_{j}\mid^{2}}{\omega_{mj}},\nonumber\\
a^s_{j}&=&\frac{\alpha_{k}E_j+i\xi E_k}{\alpha_j \alpha_k+\xi^2} ,\quad j,k=1,2,\quad j\neq k,
\end{eqnarray}
where
\begin{equation}\label{eqdet}
\alpha_j=\kappa_j+i\Delta_j
\end{equation}
and
\begin{equation}\label{7}
\Delta_j=\Delta_{0j}-\frac{g_j^{2}}{\omega_{mj}}|a^s_{j}|^2
\end{equation}
are the effective detuning. The last equation of (\ref{5}) is in fact a nonlinear equation giving the stationary intracavity field amplitude $a^s_{j}$, as the effective cavity detuning $\Delta_{j}$, comprising radiation pressure effects, is given by (\ref{7}). The parameter regime appropriate for generating optomechanical entanglement is that with a very large input power $P$, i.e., as $|a^s_{j}|\gg 1$. On the other hand, The nonlinear equation is a feature that the stationary intracavity field amplitude can reveal instability behavior for certain parameter regime. In the following, we are concerned by investigating mechanical resonator mode entanglement in the regime where the multipartite system is stable.

We proceed to  linearize the previous equations (2) about the stationary solutions by assuming that the operators are shifted by small fluctuation from their steady state solutions  $a_j = a^{s}_{j}+\delta a_j , q_j = q^{s}_{j}+\delta q_j$ and $ p_j = p^{s}_{j}+\delta p_j$.
We ignore here the nonlinear terms $\delta a_j \delta a_j^\dag$ and $\delta q_j\delta a_j$ assuming that the mean values are significantly greater than the variations. Indeed this situation can be achieved when $|a^{s}_{j}| \gg 1$. Then, we get a system of linearized quantum Langevin equations:
\begin{eqnarray}
    \delta\dot{q_{j}}&=&\omega_{mj} \ \delta p_{j}, \nonumber\\
    \delta\dot{p_{j}}&=&-\omega_{mj} \ \delta q_{j}-\gamma_{mj} \ \delta p_{j} + G_{j} \delta X_{j} +b^{in}_{j}, \nonumber\\
    \delta\dot{X_{j}}&=&-\kappa_{j} \delta X_{j} - \Delta_{j} \delta Y_{j}-\xi \delta Y_{k} + \sqrt{2\kappa_{j}}\delta X_{j}^{in},      \nonumber\\
    \delta\dot{Y_{j}}&=&-\kappa_{j} \delta Y_{j} + \Delta_{j} \delta X_{j} +\xi \delta X_{k}+ G_{j} \delta q_{j} + \sqrt{2\kappa_{j}}\delta Y_{j}^{in}.
\end{eqnarray}
where $G_{j} =\sqrt{2}g_{j} a^s_{j}$ is the effective optomechanical coupling and $a^s_{j}$ is assumed real. In addition, we have introduced the quadratures of the field mode
\begin{equation}
   \delta X_{j} = \frac{\delta a_{j} +\delta a_{j}^\dag}{\sqrt{2}},\quad \delta Y_{j} = \frac{\delta a_{j} -\delta a_{j}^\dag}{i \sqrt{2}}
\end{equation}
and the analogous Hermitian input quadrature noise amplitudes
\begin{equation}
    \delta X_{j}^{in} = \frac{a_{j}^{in} + {a_j^{in}}^{\dag}}{\sqrt{2}}, \quad  Y_{j}^{in} = \frac{ a_{j}^{in} - {a_j^{in}}^{\dag}}{i \sqrt{2}}
\end{equation}
The linearized quantum Langevin equations show that in each optomechanical cavity the mechanical mode is coupled to the cavity mode quadrature fluctuations by the effective optomechanical coupling $G_j=\sqrt{2}g_{j}a^{s}_{j}$, which can be chosen very large by increasing the intercavity amplitude $a^{s}_{j}$. On the other hand, the two cavity modes are coupled by the hopping effect process, with coupling strength $\xi$, which can be chosen such that the entanglement can be redistributed between different bipartite system and generated between the mechanical modes.

\section{Entanglement Measure}
\label{sec:3}
The mechanical and intracavity optical modes form a bipartite continuous variable (CV) system. We will explore some interesting stationary properties of this system. Since the equations (8) are linear and the noise operators are assumed to be in Gaussian state with zero-mean value, the system then can be completely categorized by its symmetrized covariance matrix (CM), which reads~\cite{35}
\begin{equation}
\nu_{lm}=\frac{\langle \mu_{l}(ss) \mu_{m}(ss)+\mu_{m}(ss)\mu_{l}(ss)\rangle}{2},
\end{equation}
where $\mu_{l}(ss)$ is the steady state value of the $l_{th}$ component of the vector of the quadrature fluctuations
\begin{equation}
\mu(t)=(\delta q_{1}(t),\delta p_{1}(t),\delta X_{1}(t),\delta Y_{1}(t),\delta q_{2}(t),\delta p_{2}(t),\delta X_{2}(t),\delta Y_{2}(t))^{T} .
\end{equation}
The equations of motion for the components of the matrix can be written as
\begin{equation}\label{sys}
    \dot{\mu}(t)=C\mu(t)+ B(t) ,
\end{equation}
where $C$ is the drift matrix

\begin{equation}\label{Dmat}
C=
\left(
 \begin{array}{cccccccc}
  0            & \omega_{m1}   & 0            & 0               & 0              & 0                & 0                  & 0              \\
  -\omega_{m1} & -\gamma_{m1}      & G_{1}               & 0              & 0            & 0                  & 0          & 0 \\
  0            & 0             & -\kappa_{1}            & \Delta_{1}     & 0              & 0                & 0                  & -\xi              \\
 G_{1}             & 0             & -\Delta_{1} & -\kappa_{1}           & 0                & 0                  & \xi          & 0 \\                                                                0            & 0             & 0            & 0               & 0 & \omega_{m2}  & 0              & 0           \\
  0       & 0             & 0            & 0               & -\omega_{m2}    & -\gamma_{m2}& G_{2}            & 0              \\
  0            & 0             & 0            & -\xi               & 0              & 0             & -\kappa_{2}  & \Delta_{2}   \\
  0            & 0             & \xi        & 0               & G_{2}             & 0                & -\Delta_{2}      & -\kappa_{2}
  \end{array}
\right),
\end{equation}
and $B(t)$ is the vector composed of noise terms
\begin{equation}
B=(0,b_{1}^{in}(t),\sqrt{2\kappa_{1}}X_{1}^{in}(t),\sqrt{2\kappa_{1}}Y_{1}^{in}(t),0, b_{2}^{in}(t),\sqrt{2\kappa_{2}}X_{2}^{in}(t),\sqrt{2\kappa_{2}}Y_{2}^{in}(t))^{T}.
\end{equation}
We can show that in the absence of the hopping process, we recover the form of the two independent optomechanical cavities, which are represented by diagonal blocks of the drift matrix. Furthermore, in this case the entanglement is confined in each intracavity. However, one can seen from the diffusion matrix that the cavity modes decay to a common reservoir, so the entanglement can be transferred between the two independent optomechanical cavities. On the other hand, the entanglement transfer depends on the environment. When the hopping process is introduced that  the non-diagonal blocks become nonzero, the entanglement redistribution between the mechanical modes can exist without the coupling to the environment. In the following, we explore this point in details.

The steady state CM can be determined by solving the Lyapunov equation \cite{35}
\begin{equation}
    CZ+ZC^T=-\mathcal{D},
\end{equation}
where $\mathcal{D}$ represents the diffusion matrix , which is determined by the noise correlation functions
\begin{equation}
\mathcal{D}=diag \Big(0,\gamma_{m1}(2\bar{n}+1),\kappa_{1},\kappa_{1},0,\gamma_{m2}(2\bar{n}+1),\kappa_{2},\kappa_{2}\Big)
\end{equation}
The CM allows to calculate the stationary entanglement. Indeed, to calculate the pairwise entanglement, we reduce the $(8 \times 8)$ covariance matrix $Z$ to a $(4 \times 4)$ submatrix $Z_R$. There are four such cases of the submatrix $Z_R$: (i) if the indices $i$ and $j$ for the element $z_{ij}$ are confined to the set $\big\{1, 2, 3, 4\big\}$, the submatrix $Z_R=[z_{ij}]$ is produced by the first four rows and columns of $Z$ and correspond to the covariance between the first intracavity photon-phonon coupling. Similarly, (ii) if the indices run over $\big\{5, 6, 7, 8\big\}$, $Z_R$ is the covariance matrix of the second intracavity photon-phonon interaction. (iii) If the indices run over $\big\{1, 2, 5, 6\big\}$, $Z_R$ designates the covariance between the two mechanical resonator modes. (iv) Finally, if the indices run over $\big\{3, 4, 7, 8\big\}$, $Z_R$ represents the covariance matrix between the cavity modes.

Thus, the logarithmic negativity $EN$ \cite{36} can be adopted here as a good entanglement measure, and can be written as \cite{35}
\begin{equation}
    EN= \max\Big[0,-\ln(2\vartheta^{-})\Big],
\end{equation}
where
\begin{equation}
    \vartheta^{-}= \frac{1}{\sqrt{2}}\big(\chi(Z_R)-[\chi(Z_R)^2-4\det(Z_R)]^{1/2}\big)^{1/2},
\end{equation}
and $\chi(Z_R)\equiv \det Z_{1}+\det Z_{2}-2\det Z_{c}$, with $Z_{1}, Z_{2}$ and $Z_{c}$ being 2$\times$ 2 block matrices
\begin{equation}
Z_R =\left(
 \begin{array}{cc} Z_{1} & Z_{c}  \\Z_{c}^{T} & Z_{2}\end{array}
\right).
\end{equation}
Clearly, we can see that the necessary and sufficient condition for the Gaussian state being entangled can be read as
\begin{equation}
  \vartheta^{-} < \frac{1}{2},
\end{equation}
which is entirely  identical  to the Simon's criterion which states that the necessary and sufficient condition for entanglement of non-positive partial transpose condition for Gaussian states \cite{37} is given by
\begin{equation}
    4\det Z_R < \chi(Z_R)-\frac{1}{4}.
\end{equation}
The logarithmic negativity is a good and suitable measure for entanglement. Furthermore,  it can always be explicitly calculated. It has an important feature that is additive measure.  On the other hand, the steady state regime of the system is Gaussian state as the condition of the linearization method is valid. Within this context, we are now ready to investigate the generation of entanglement between mechanical modes.

\section{Experimental Consideration}
\label{sec:4}
As mentioned above, the model described here could be realized experimentally in the system composed of two Fabry-Perot cavities and whispering cavities \cite{32}. We explore parameter ranges in which the expected entanglement effects could be observed with the current experiments. The parameter regime very close to the current experimental results \cite{38,39,40} is given as \Big($L_{j} \simeq 1 mm, m_{j} \simeq 10 ng , \omega_{mj}/2\pi = 10 MHz, \gamma_{mj}/2\pi \simeq 100 Hz, \kappa_{j}/2\pi\simeq 5-15 MHz , T =0.6\sim 20 K, P = 50 mW , \lambda_{j} =1064 nm$\Big).
Within these values we will show that the entanglement can be generated between the two mechanical resonators with the proposed scheme.
For simplicity, without ignoring the generality,  we suppose that the two cavities are identical and we choose the same parameters for the two mechanical resonators and the two cavity modes, i.e; $\omega_{m1}=\omega_{m2}=\omega_m, \kappa_1=\kappa_2=\kappa, \gamma_{m1}=\gamma_{m2}=\gamma_m, T_1=T_2=T, G_1=G_2=G, \Delta_{1}=\Delta_{2}=\Delta$

\section{ Results and Discussion}
\label{sec:5}
\begin{figure}[h]
 \center{\includegraphics[width=0.4\linewidth,height=0.33\linewidth]{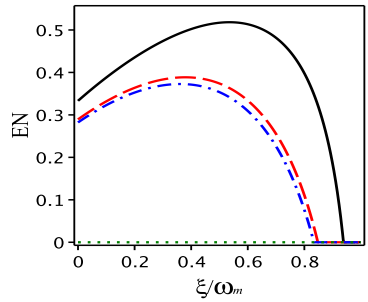}}
\caption{Logarithmic negativity $EN$ of mechanical modes as a function of the normalized photon-hopping coupling strength $\xi/\omega_{m}$ , with $\Delta/\omega_m=1 $, $\gamma_m/\omega_m = 10^{-5}$, $\kappa/\omega_m=0.5$, $T=0.6 K$ and for different values of the normalized effective coupling strength $G/\omega_m$: dotted green line $G/\omega_m=1$, solid black line $G/\omega_m=4$,
dashed red line $G/\omega_m=8$ and dash dotted blue line $G/\omega_m=12$}\label{Fig2}
\end{figure}

First, we investigate the ranges of the parameters that produce the stability conditions within the experimental considerations. As it has previously stated the system (\ref{sys}) is stable and attained its steady state when the entire eigenvalues of the drift matrix (\ref{Dmat}) have negative real parts. On the other hand the stability conditions can be derived using the Routh-Hurwitz criterion \cite{41}. Since the drift matrix is of order $(8 \times 8)$, the expressions are more complex and cannot be presented here. Without lost of generality, we explore the numerical calculations of the logarithmic negativity for different parameter values within the experimental domain. In figure \ref{Fig2}, we plot the stationary logarithmic negativity $EN$ in terms of the normalized photon-hopping coupling strength $\xi/ \omega_{m}$ for $\Delta/\omega_m=1, \kappa/\omega_m=0.5, \gamma_m/\omega_m=10^{-5}$  and for different values of the normalized effective coupling strength $G/\omega_m$. We can clearly see that below threshold of the effective coupling strength there is no entanglement generation between mechanical modes. Moreover, above the threshold the logarithmic negativity maintains the same behavior and its amount decreases with increasing the normalized effective coupling strength $G/\omega_m$. On the other hand the effect of the photon hopping process increases the value of the logarithmic negativity indicating the presence of a stationary correlated state between mechanical modes. The logarithmic negativity reaches its maximum value around $\xi/\omega_m \simeq 0.5$, which agrees with the recent results \cite{31}. It then decreases to zero when $\xi/\omega_m \gtrsim 1$

\begin{figure}[h]
\center{\includegraphics[width=0.4\linewidth,height=0.33\linewidth]{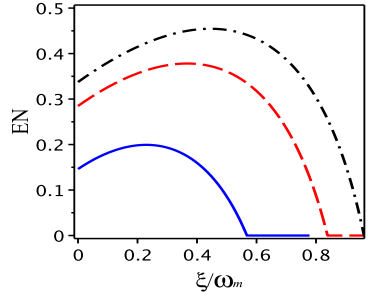}}
\caption{Logarithmic negativity $EN$ of mechanical modes as a function of the normalized photon-hopping coupling strength $\xi/\omega_{m}$ , with $G/\omega_m=10$, $\gamma_m/\omega_m = 10^{-5}$, $\kappa/\omega_m=0.5$, $T=0.6 K$ and for different values of normalized detuning: solid blue line $\Delta/ \omega_{m}= 0.8$,
dashed red line $\Delta/ \omega_{m}= 1.0$, dash dotted black line $\Delta/ \omega_{m}= 1.1$}\label{Fig3}
\end{figure}

In order to explore the dependence of the generated entanglement on the photon hopping process and the effective detuning, we plot in figure \ref{Fig3} the logarithmic negativity between the mechanical modes versus the normalized photon hopping coupling strength $\xi/ \omega_{m}$ for different values of the normalized effective detuning. As can be seen from the figure, in order to achieve maximum entanglement for a given effective detuning one has to apply a certain photon hopping process. However, we find that there exists an optimum amount of photon hopping effect that is needed to obtain the maximum entanglement for the realistic set of parameters. On the other hand, the amount of the logarithmic negativity increases with increasing the effective detuning values within the experimental ranges and the range of the effective detuning values are also governed by the stability conditions.

\begin{figure}[h]
%\begin{flushleft}\quad\quad\quad\textbf{(a)}\end{flushleft}
\hspace*{2cm}\textbf{(a)} \hspace*{3.5cm}\textbf{(b)}\hspace*{3.5cm}\textbf{(c)}\\
\includegraphics[width=0.33\linewidth,height=0.33\linewidth]{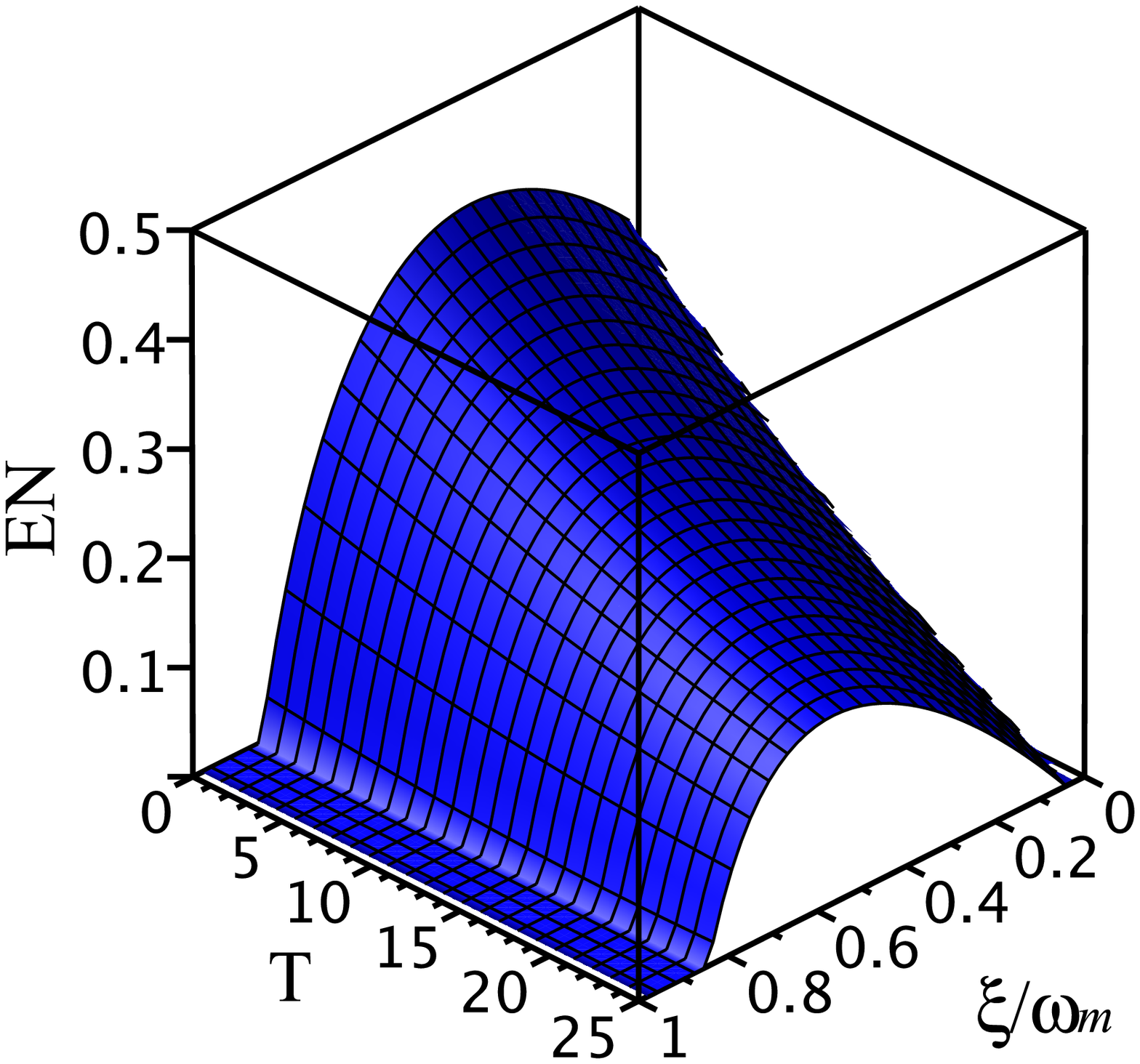}
\includegraphics[width=0.33\linewidth,height=0.33\linewidth]{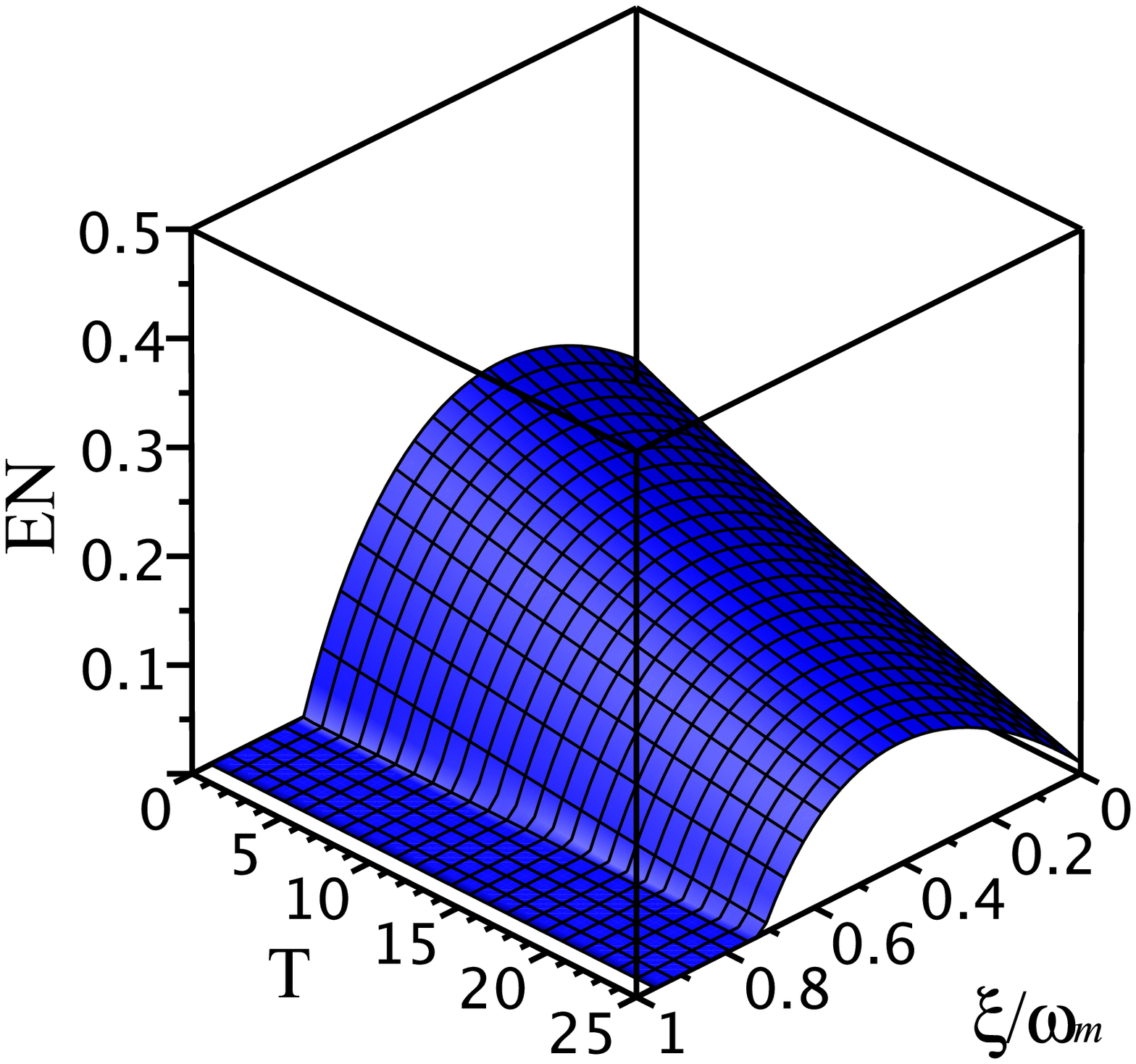}
\includegraphics[width=0.33\linewidth,height=0.33\linewidth]{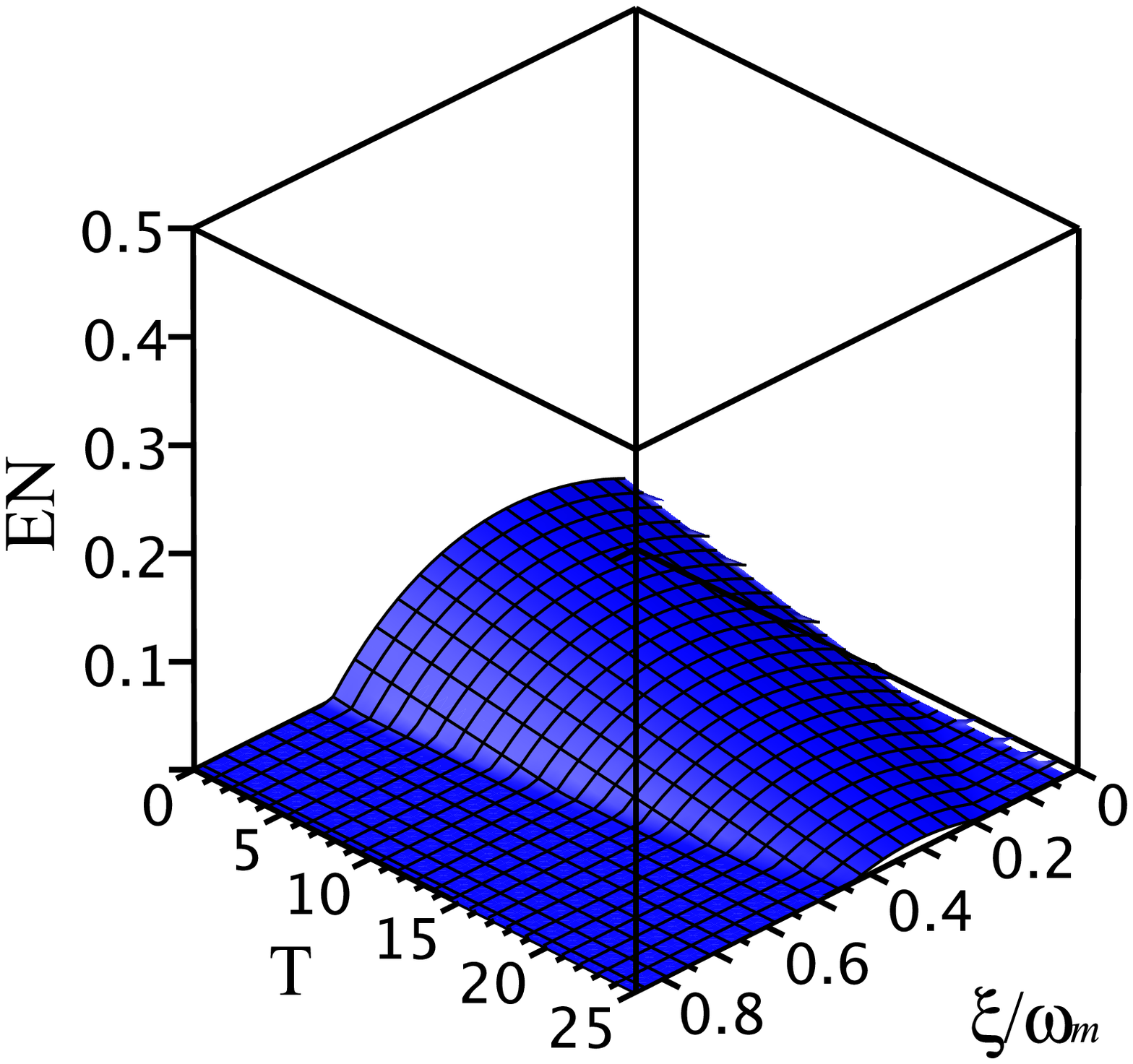}
\caption{(Color online) Stationary logarithmic negativity $EN$ of the the mechanical modes in terms of the normalized photon-hopping coupling strength $\xi/ \omega_{mj}$ and the environment temperature $T$  for normalized detuning $\Delta/ \omega_{m}=1 $, the effective optomechanical coupling $G/\omega_m=8$ and the mechanical damping rate $\gamma_m/\omega_m = 10^{-5}$  for different values of the optical decay rate: (a) $\kappa/\omega_m=0.5$, (b) $\kappa/\omega_m=1$ and (c) $\kappa/\omega_m=1.5$.  }
\label{Fig4}
\end{figure}

Finally, we examine the variation of the entanglement between the mechanical modes versus the photon hopping strength and the temperature of the phonon reservoir. Figure \ref{Fig4} shows the variation of the logarithmic negativity $EN$ as a function of the normalized photon hopping $\xi/\omega_m$ and the temperature of the phonon reservoir $T$ for the normalized detuning $\Delta/ \omega_{m}=1$, $\gamma_m/\omega_m=10^{-5}$, $G/\omega_m=8$ and different values of the optical decay rate $\kappa/\omega_m =0.5, 1.0, 1.5$. We see that optomechanical entanglement is generated just in a restricted range of values of $\xi$ around $\xi/\omega_m\sim 0.5$. The strength of the entanglement with respect to the temperature can be also inspected. The appropriate results of mechanical mode entanglement persevere for temperatures above $20K$, which is some orders of magnitude larger than the ground state temperature of the mechanical resonators. When the optical decay rate augments the entanglement decreases as it has shown in the figures (4a, 4b, ac). We have found that the logarithmic negativity is present even for $\kappa/\omega_m > 1$, although it becomes much less strong against temperature.

%\clearpage
\section{Conclusions}
\label{sec:7}
We have investigated a standard scheme for exploring the optomechanical entanglement. We have examined the technique of continuous variables to investigate the main characteristic of quantum mechanics that is called entanglement.
On the other hand, the standard scheme illustrates that the entanglement can be generated between the mechanical resonators by introducing the photon hopping interaction between two cavity modes.  In addition, we have use the logarithmic negativity as a good measure to quantify the amount of entanglement between bipartite systems.
We have shown that the amount of stationary entanglement between the mechanical resonators increase with increasing the power of laser that use to derive the two cavities. Further, this amount of the stationary entanglement is comparable to that of the Bell states. Moreover, the used parameters are experimentally viable. This may advantage forward attaining quantum ground state of mechanical resonators in experiments and advance possible applications concerning quantum information processing supported on mechanical resonators. The implementation of this sort of entanglement at the new macroscopic rank of micromechanical resonators can be exceptionally central both for practical and fundamental motives. In reality, on the one hand, entangled two spatially separated mechanical resonator could characterize a significant structure block for the performance of quantum networks for long-distance routing of quantum information; on the other hand, these nonclassical states signify a perfect platform for examining and comparing decoherence theories and modifications of quantum mechanics at the macroscopic level.
The extension to other much complicated hybrid systems comprise a motivating field of research. Furthermore, the recognition of these schemes will open original viewpoint for the perceptive of quantum memories for continuous variable (CV) technique in quantum information processing.

%\begin{acknowledgements}
%\end{acknowledgements}



\end{document}